\documentclass{aastex631}

\usepackage{amsmath}
\usepackage{subfigure}
\usepackage{textcomp}
\usepackage[misc]{ifsym} %% Letter symbol
\usepackage{hyperref}
\usepackage{threeparttable}
\usepackage{CJK}
\shortauthors{Tao et al.}
\graphicspath{{./}{figures/}}
\begin{document}
\begin{CJK*}{UTF8}{gbsn}
\title{The most sensitive SETI observations toward Barnard's star with FAST}

\author[0000-0002-4683-5500]{Zhen-Zhao Tao}
%\altaffiliation{These authors contributed equally to this work.}
\affiliation{Institute for Frontiers in Astronomy and Astrophysics, Beijing Normal University, Beijing 102206, China}
\affiliation{Department of Astronomy, Beijing Normal University, Beijing 100875, China; \url{tjzhang@bnu.edu.cn}}
\affiliation{Institute for Astronomical Science, Dezhou University, Dezhou 253023, China}

\author[0000-0002-8719-3137]{Bo-Lun Huang}
%\altaffiliation{These authors contributed equally to this work.}
\affiliation{Institute for Frontiers in Astronomy and Astrophysics, Beijing Normal University, Beijing 102206, China}
\affiliation{Department of Astronomy, Beijing Normal University, Beijing 100875, China; \url{tjzhang@bnu.edu.cn}}

\author[0000-0003-3977-4276]{Xiao-Hang Luan}
\affiliation{Institute for Frontiers in Astronomy and Astrophysics, Beijing Normal University, Beijing 102206, China}
\affiliation{Department of Astronomy, Beijing Normal University, Beijing 100875, China; \url{tjzhang@bnu.edu.cn}}

\author[0000-0002-5485-1877]{Jian-Kang Li}
\affiliation{Institute for Frontiers in Astronomy and Astrophysics, Beijing Normal University, Beijing 102206, China}
\affiliation{Department of Astronomy, Beijing Normal University, Beijing 100875, China; \url{tjzhang@bnu.edu.cn}}

\author[0000-0002-5485-1877]{Hai-Chen Zhao}
%\altaffiliation{These authors contributed equally to this work.}
\affiliation{Institute for Frontiers in Astronomy and Astrophysics, Beijing Normal University, Beijing 102206, China}
\affiliation{Department of Astronomy, Beijing Normal University, Beijing 100875, China; \url{tjzhang@bnu.edu.cn}}

%\collaboration{6}{(AAS Journals Data Editors)}
\author{Hong-Feng Wang}
\affiliation{Institute for Astronomical Science, Dezhou University, Dezhou 253023, China}

\author[0000-0002-3363-9965]{Tong-Jie Zhang(张同杰)\href{mailto:tjzhang@bnu.edu.cn}{\textrm{\Letter}}}
\affiliation{Institute for Frontiers in Astronomy and Astrophysics, Beijing Normal University, Beijing 102206, China}
\affiliation{Department of Astronomy, Beijing Normal University, Beijing 100875, China; \url{tjzhang@bnu.edu.cn}}
\affiliation{Institute for Astronomical Science, Dezhou University, Dezhou 253023, China}

\begin{abstract}

Search for extraterrestrial intelligence (SETI) has been mainly focused on nearby stars and their planets in recent years. Barnard's star is the second closest star system to the sun and the closest star in the FAST observable sky which makes the minimum Equivalent Isotropic Radiated Power (EIRP) required for a hypothetical radio transmitter from Barnard's star to be detected by FAST telescope a mere $4.36 \times 10^{8} ~\text{W}$. In this paper, we present the Five-hundred-meter Aperture Spherical radio Telescope (FAST) telescope as the most sensitive instrument for radio SETI observations toward nearby star systems and conduct a series of observations to Barnard's star (GJ~699). By applying the multi-beam coincidence matching (MBCM) strategy on the FAST telescope, we search for narrow-band signals ($\sim$Hz) in the frequency range of 1.05-1.45 GHz, and two orthogonal linear polarization directions are recorded. Despite finding no evidence of radio technosignatures in our series of observations, we have developed predictions regarding the hypothetical extraterrestrial intelligence (ETI) signal originating from Barnard's star. These predictions are based on the star's physical properties and our observation strategy.

\end{abstract}

\keywords{\href{http://astrothesaurus.org/uat/74}{Astrobiology (74)}; \href{http://astrothesaurus.org/uat/2127}{Search for extraterrestrial intelligence (2127)}; \href{http://astrothesaurus.org/uat/2128}{Technosignatures (2128)}; \href{http://astrothesaurus.org/uat/498}{Exoplanets (498)}}

%% We recommend that authors also use the natbib  \citeppp
%% and  \citept commands to identify citations.  The citations are

\section{Introduction} \label{sec:intro}

%SETI is a multidisciplinary scientific endeavor that seeks to answer one of humanity's most profound questions: Are we alone in the universe\citep{2001ARA&A..39..511T}?
The Search for Extraterrestrial Intelligence (SETI) is a scientific endeavor aimed at detecting and studying potential signals from intelligent civilizations beyond Earth. The significance of SETI lies in its potential to answer one of the most profound questions in human history: Are we alone in the universe? 
Since the pioneering work of Frank Drake and his eponymous equation in the early 1960s \citep{1961PhT....14d..40D}, SETI has evolved into a diverse field encompassing radio and optical astronomy, astrobiology, and even the development of advanced technologies for interstellar communication \citep{2011AcAau..68..347S}. The discovery of thousands of exoplanets orbiting stars beyond our solar system has further invigorated the search for extraterrestrial life, as these distant worlds may harbor conditions suitable for the emergence of life as we know it \citep{2013Sci...340..577S}.

Narrow-band radio SETI started as early as the 1960s \citep{1961PhT....14d..40D} because of several advantages. The natural and thermal broadening effect that makes signals produced by astrophysical sources possess a minimum width of hundreds of hertz \citep{1997ApJ...487..782C}. Narrowband signals, on the other hand, are necessarily technologically modulated. As a result, narrowband signals can serve as technosignature of ETI and can be distinguished from natural interference in the radio sky. Therefore, narrowband signals are the most searched type of signals in radio SETI \citep{2013ApJ...767...94S,2017ApJ...849..104E,2020AJ....159...86P,2020ApJ...895L...6Z,2022AJ....164..160T,2019AJ....157..122P,2016AJ....152..181H,2021AJ....162...33G,2021AJ....161..286T}.

The Five-hundred meter Aperture Spherical radio Telescope (FAST) is a groundbreaking instrument that has revolutionized our ability to explore the universe \citep{2011IJMPD..20..989N}.
Located in the Guizhou Province of China, FAST is the world's largest single-dish radio telescope. Its innovative design, which features an active reflector surface and a feed cabin suspension system, allows for unparalleled sensitivity and flexibility in observing a wide range of celestial phenomena \citep{2019SCPMA..6259502J}.
Since its installation in 2016, 
%FAST has played a crucial role in advancing SETI research by detecting and analyzing radio signals from distant stars and galaxies
FAST has played a crucial role in searching for radio signals from distant stars and galaxies
\citep{2020ApJ...891..174Z,2022AJ....164..160T,2022ApJ...938....1L,2023AJ....165..132L}. Its exceptional capabilities have enabled scientists to search for potential signs of ETI. Furthermore, the sensitivity of FAST is perfectly suitable for searching technosignatures from nearby stars. For an ETI transmitter inside Ross 128 system, the minimum Equivalent Isotropic Radiated Power (EIRP) that is required for FAST to detect is $1.48 \times 10^{9} ~\text{W}$ , which is even a few orders of magnitude less than the EIRP of the Arecibo Planetary Radar of $\sim$10$^{13}$~W \citep{2022AJ....164..160T}.

Barnard's Star is a red dwarf located in the constellation of Ophiuchus, approximately 1.8 parsecs away from our solar system \citep{2023A&A...674A...1G}. Named after the American astronomer Edward Emerson Barnard, who first observed its high proper motion in 1916 \citep{1916AJ.....29..181B}, this star has since become a subject of extensive study and speculation.
As the second closest red dwarf to our solar system, after Proxima Centauri, Barnard's Star has garnered attention for its potential to host exoplanets and the possibility of harboring life \citep{2016Natur.536..437A}. In recent years, the discovery of a candidate exoplanet, Barnard's Star b, has further fueled scientific curiosity and investigation. This super-Earth, with a minimum mass of 3.2 Earth masses, the maximum equilibrium temperature of the planet is 105 K
%orbits its host star in a frigid environment
, raising questions about its habitability and the potential existence of other planets in the system \citep{2018Natur.563..365R}, however \cite{2021AJ....162...61L}demonstrates that the planet candidate is most likely a false positive.
     
In this paper, we present the results of the SETI observations of Barnard's Star. In Section \ref{sec:Obser}, we introduce the multi-beam coincidence matching (MBCM) strategy and the data collecting pipeline. The techniques we use to analyze our data are discussed in Section \ref{sec:DataAnalysis}. The results are presented in Section \ref{sec:Results}. In Section \ref{sec: Discussion}, we discuss the sensitivity of our observations and a simple simulation of the ETI signal originating from Barnard's Star b. Finally, the conclusions of this work are presented in Section \ref{sec: Conclusions}.

%This paper is organized as follows. In section 2, the observation strategy and data analysis method are given. The results are given in section 3, followed by discussion and conclusions.

\section{Observations} \label{sec:Obser}
Identifying man-made radio frequency interference (RFI) has always been a major challenge in radio SETI, particularly in an era of increasing radio pollution. The ON-OFF strategy is employed in target observations to effectively identify RFI \citep{2013ApJ...767...94S,2017ApJ...849..104E,2019AJ....157..122P,2020AJ....159...86P,2020AJ....160...29S,2021AJ....161..286T,2021AJ....162...33G,2021NatAs.tmp..203S}, where intermittent observations of a reference position (OFF observation) are interleaved with observations of the target source (ON observation). The angular separation between the ON and OFF sources needs to be at least several times the telescope’s half-power beamwidth (HPBW) to ensure that a sky-localized signal from the ON source is not detected during the OFF observations. 

RFI signals are mostly not sky-localized, so they can be visible in both ON and OFF observations and thus can be directly eliminated. The MBCM strategy follows the same principle by taking the ON and OFF observation simultaneously with the FAST L-band 19-beam receiver \citep{2022AJ....164..160T,2023AJ....165..132L}. Only Beam 1 keeps tracking right on the target star, acting as the ON observation, whereas the six outermost beams (Beam 8, 10, 12, 14, 16, and 18) play the roles of the OFF observations. We refer to the narrowband signals detected above a certain signal-to-noise ratio (S/N) threshold by any of the 19 beams as hits. Hits that possess an S/N greater than the threshold in the center beam but below the threshold in all the reference beams are called events. The signals detected by both Beam 1 and any of the reference beams are determined as RFI and rejected immediately. Using as many as six reference beams, we are able to remove the vast majority of signals detected by Beam 1.

%\vspace{\baselineskip}
%(image of 19 beam)
%\vspace{\baselineskip}
 \begin{figure}[htbp]
 \centering
 \includegraphics[width=0.3\textwidth]{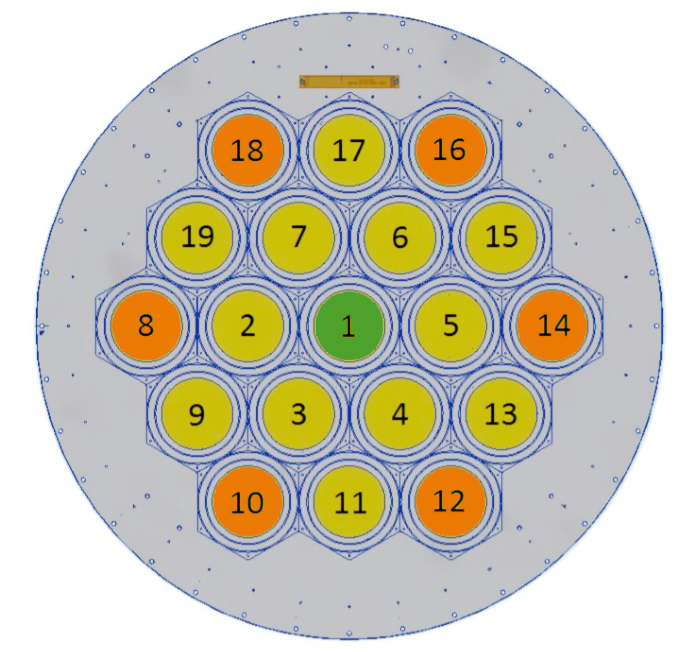}
 \caption{\label{fig:snr} Schematic diagram of FAST's 19-beam receiver, the beam filled with green is Beam 1 which tracks the target star during the entire observation; the six outermost beams filled with orange are sufficiently separated from Beam 1 and act as the reference beams. }
 \end{figure}

On July 9th, 10th, 21st, and 28th 2022, we conducted four individual observations of Barnard's Star with FAST, each with a duration of 20 minutes. We collected data in the frequency range of 1.0 - 1.5 GHz on the SETI back end, the number of frequency channels is 65536k, with a frequency resolution of approximately 7.5 Hz, and an integration time of 10 seconds for each spectrum. Each FITS file\citep{1981A&AS...44..363W} contained data from four polarized channels (XX, YY, XY, YX) of two spectra per beam. The top and bottom 50 MHz are not within the designed frequency band of the receiver. Therefore, the effective data consists of a total bandwidth of 400 MHz, ranging from 1050 MHz to 1450 MHz.

\section{Data Analysis} \label{sec:DataAnalysis}

There are currently two operating modes for the MBCM strategy: the standard targeted search mode which specifically focuses on a target star, and a blind search mode which extends the search area to the vicinity of the target star \citep{2023AJ....165..132L}. Since this work is solely dedicated to Barnard's star, we choose the standard targeted mode rather than the blind search mode.
 
A narrowband signal transmitted from a distant source drifts in frequency due to the Doppler effect, and the drift rate is given by $\dot{\nu} = \nu_{0}a/c$, where $\nu_{0}$ is the original frequency 
of the signal, $a$ is the relative acceleration between the transmitter and the receiver, and $c$ is the speed of light. Thus Doppler drift represents a search parameter for narrowband SETI.
Following \cite{2022AJ....164..160T}, we searched artificial narrowband signal data for two orthogonal linear polarization directions (XX and YY) respectively, using turboSETI \citep{2017ApJ...849..104E,2019ascl.soft06006E}. We convert the fits file obtained from observations into a filterbank file accessible by turboSETI. TurboSETI is a Python/Cython package using the tree search algorithm \citep{2013ApJ...767...94S} to search for narrow-band signals.  Two essential parameters required by turboSETI are an S/N threshold and a maximum drift rate (MDR), which allow turboSETI to search for narrow-band signals above the S/N threshold within $\pm$MDR.

It is essential to choose an appropriate value for the S/N threshold, setting an excessively high S/N can potentially filter out most of the received signals, which also increases the chance of filtering out an authentic ETI signal, whereas an insufficient S/N threshold can fail to filter out even some of the low S/N noises.

Since the drift rate of a signal is proportional to its original frequency, the suitable MDR depends on the observing frequency band, and 4 Hz/s is large enough for FAST L-band. Following \cite{2022AJ....164..160T}, we set the S/N threshold to 10 and the MDR to 4 Hz/s. TurboSETI outputs the best-fit frequencies, drift rates, and S/Ns of the hits detected by each beam to a DAT file.
 
We use the modified find\_event\_pipeline method of TurboSETI to search for hits that exist in beam 1, but not in the reference beams\citep{2022AJ....164..160T}. TurboSETI outputs the selected events of each source to a CSV file. Since the effective bandwidth of the L-band receiver is $1.05 - 1.45$ GHz \citep{2011IJMPD..20..989N}, we discard events detected within the two unusable 50 MHz-wide ends of the band after event finding by turboSETI.

\section{Results} \label{sec:Results}

%\subsection{Results of Signal Search}

\begin{figure}[htbp]
\centering
\subfigure[]{
\includegraphics[width=0.8\linewidth]{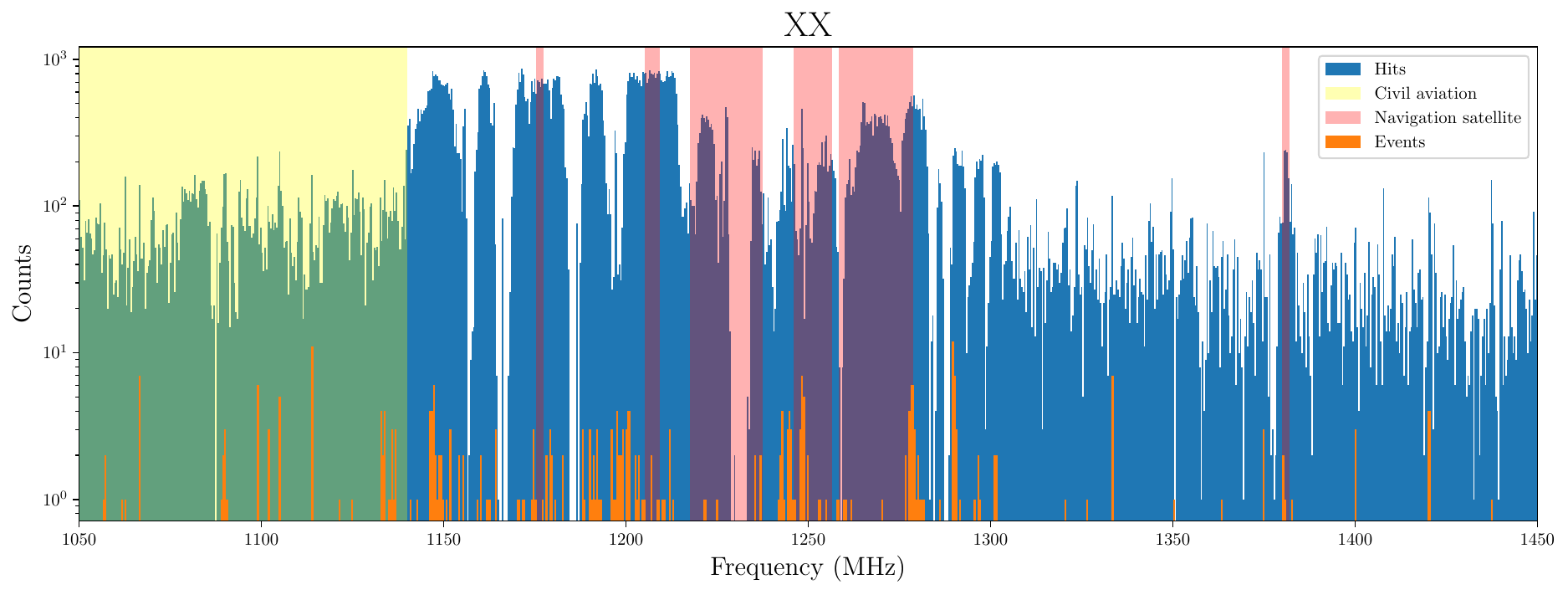}
}

\subfigure[]{
\includegraphics[width=0.4\linewidth]{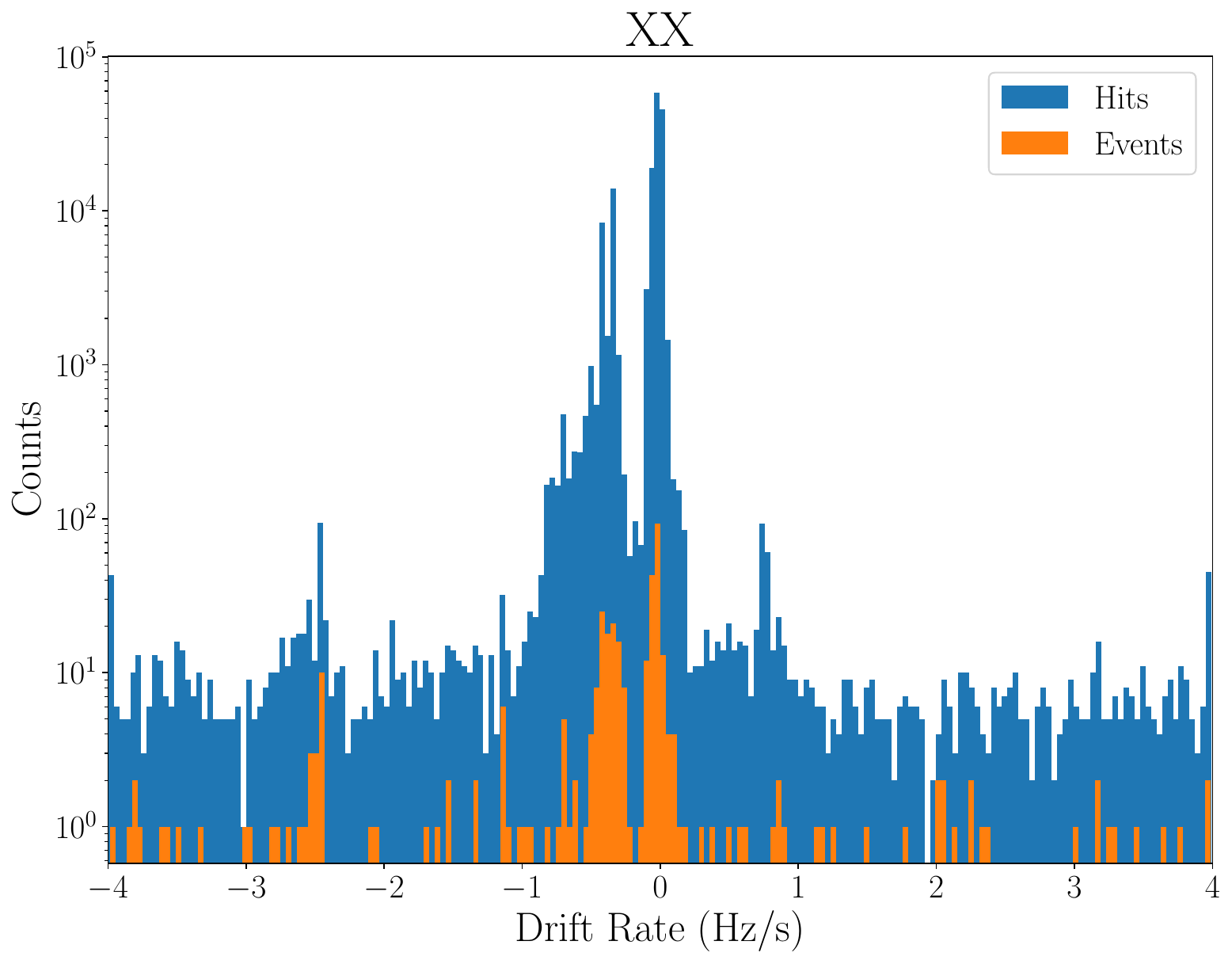}
}
\subfigure[]{
\includegraphics[width=0.4\linewidth]{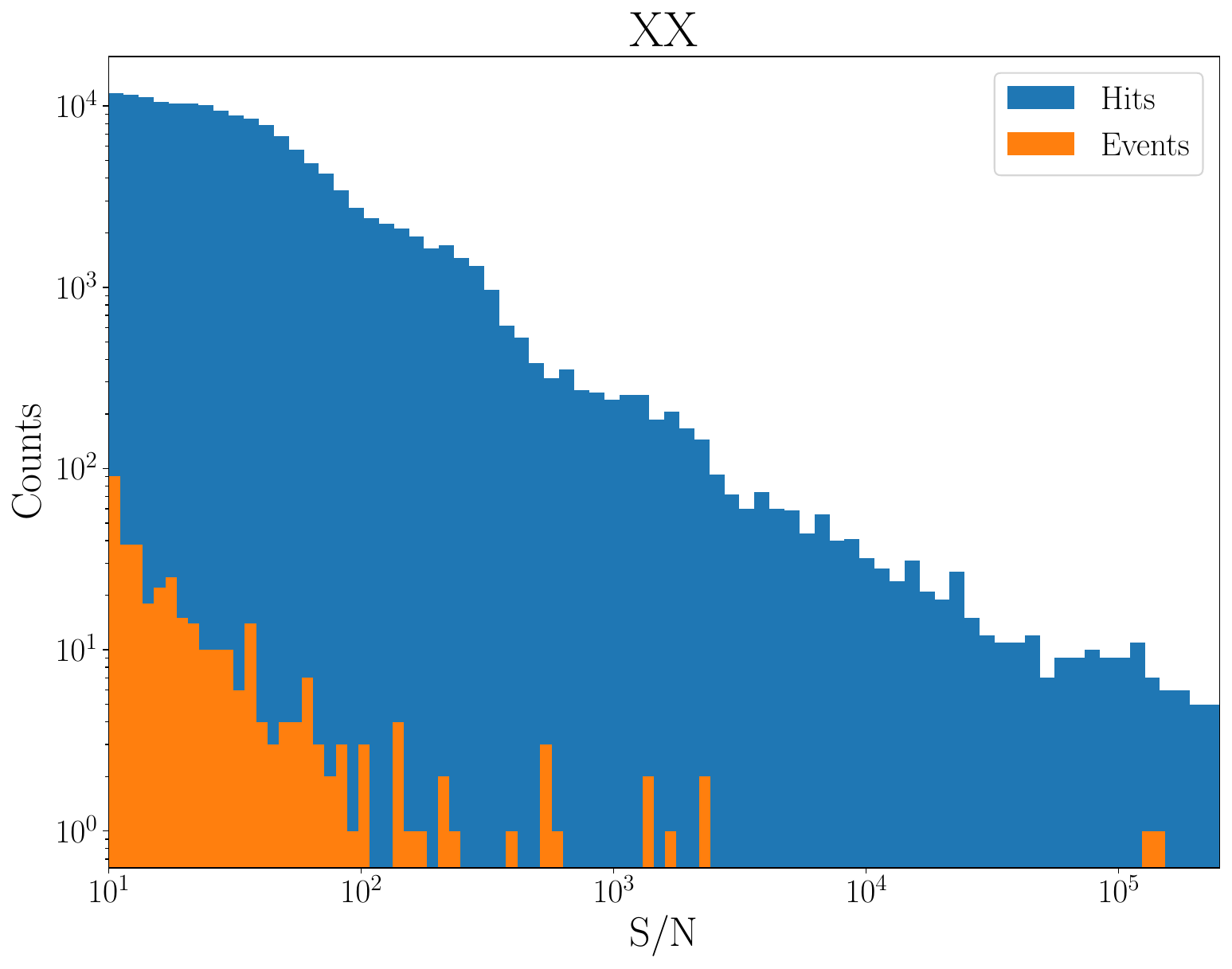}
}
\caption{Histograms of hits and events for XX as functions of frequency, drift rate and S/N. Frequency bands of catalogued interference sources are displayed on the frequency panel. The blue bars show the distributions of the hits detected by all 19 beams. The orange bars show the distributions of the events detected by Beam 1.}
\label{fig:statistics1}
\end{figure}

\begin{figure}[htbp]
\centering
\subfigure[]{
\includegraphics[width=0.8\linewidth]{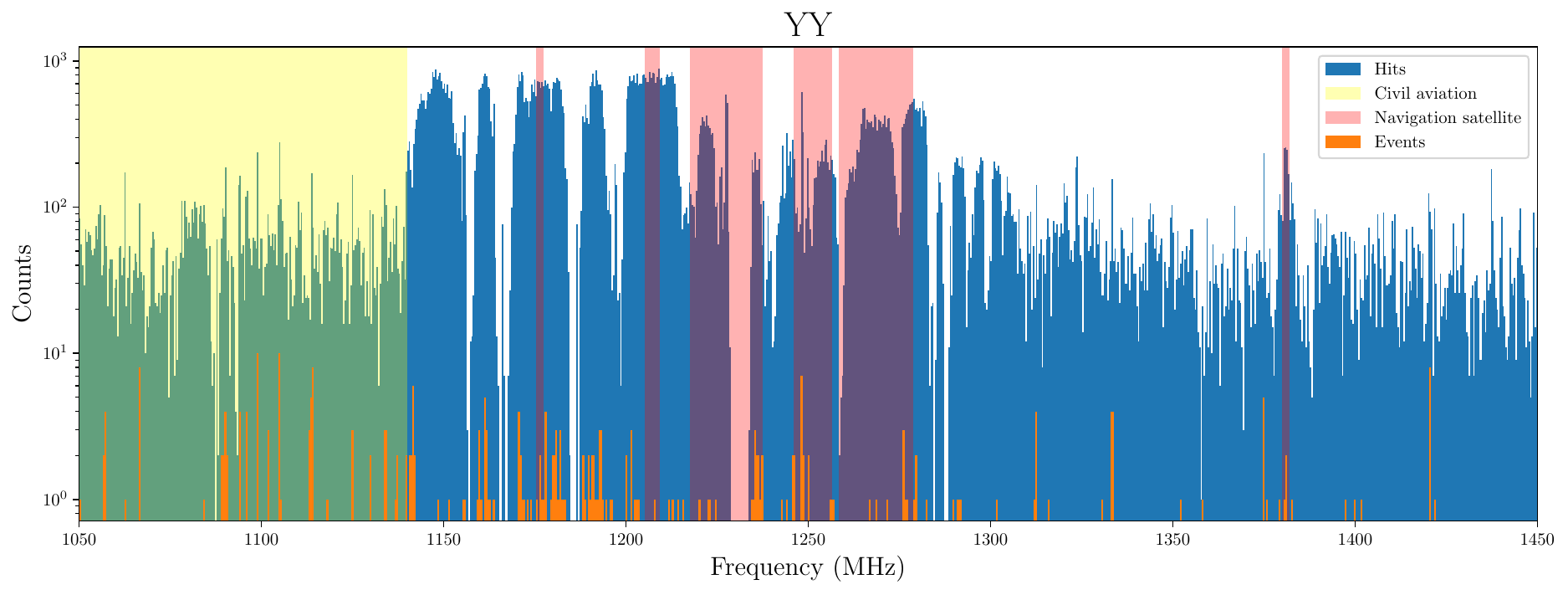}
}

\subfigure[]{
\includegraphics[width=0.4\linewidth]{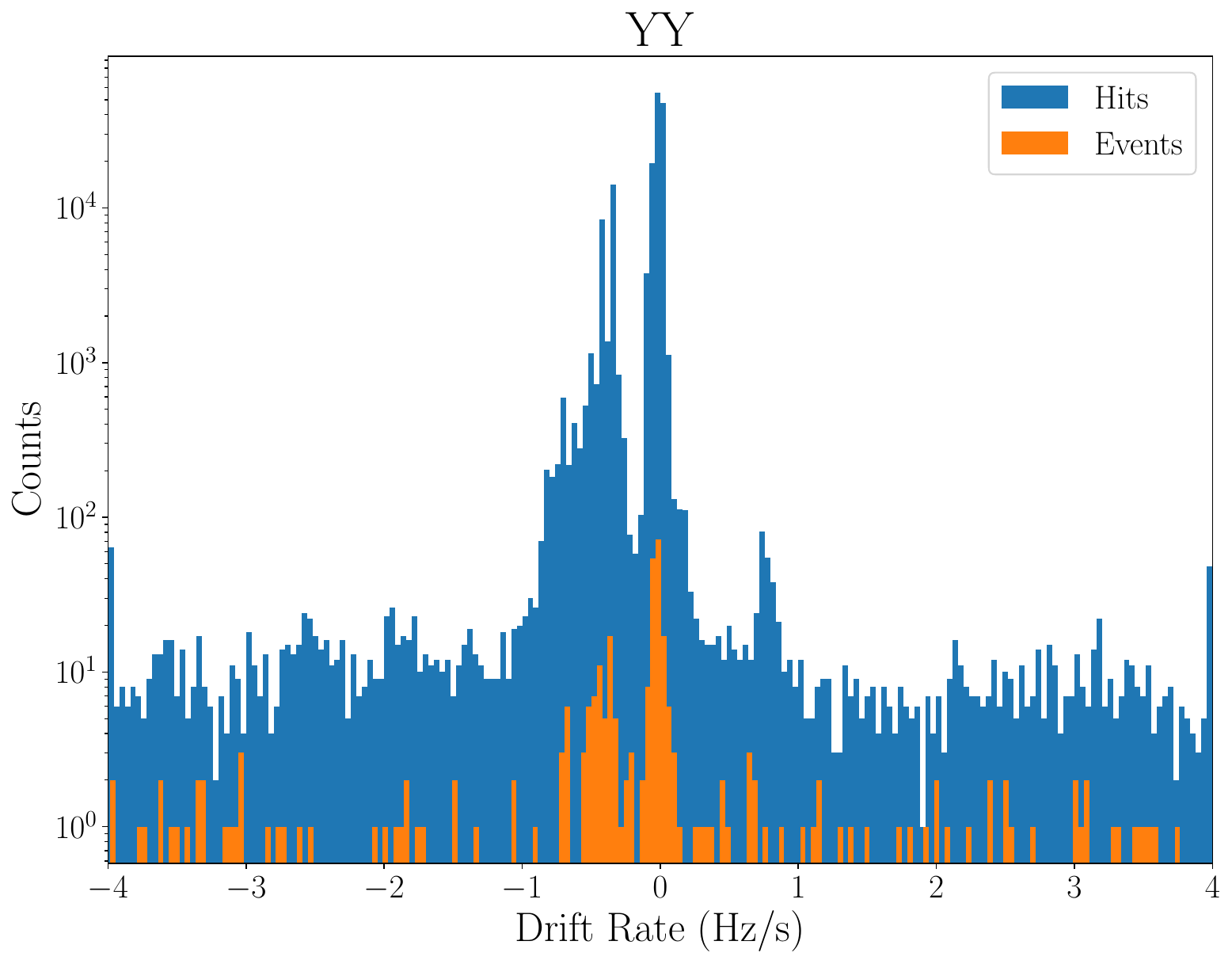}
}
\subfigure[]{
\includegraphics[width=0.4\linewidth]{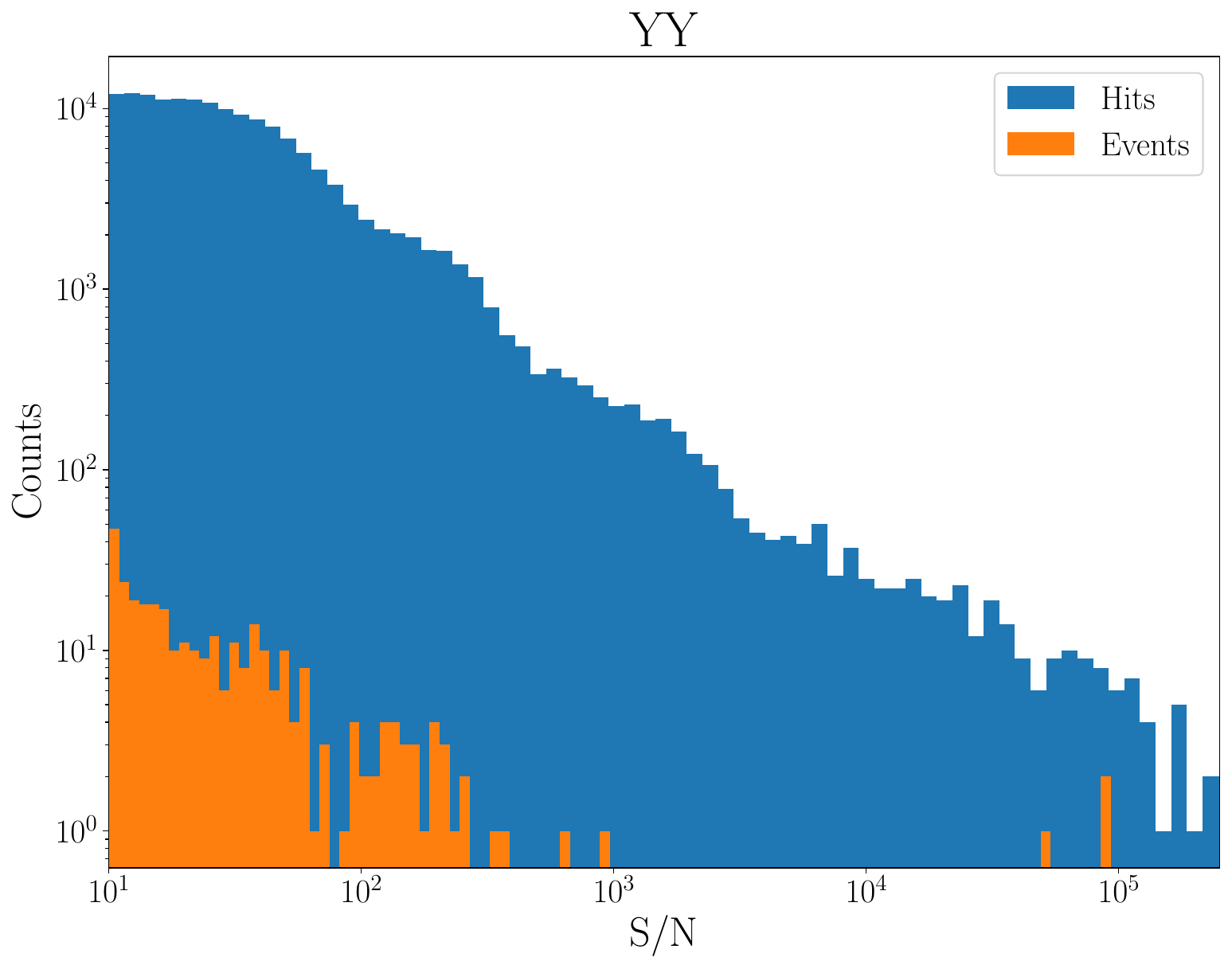}
}
\caption{Histograms of hits and events for YY as functions of frequency, drift rate and S/N.}
\label{fig:statistics2}
\end{figure}

We ran turboSETI on all observation data, finding 159,394 hits of XX polarization and 160,097 hits of YY polarization from all 19 beams. Of these, 367 events of XX and 317 events of YY were detected only in Beam 1.
%, excluding 97.0\% and 96.9\% of the hits detected by Beam 1, respectively.
The distributions of frequency, drift rate, and S/N for the hits detected by all 19 beams and the events detected by Beam 1 are shown in Figure \ref{fig:statistics1} and \ref{fig:statistics2}, and each figure displays one polarization direction. The general distribution trend of each parameter does not vary apparently with polarization direction, implying that most hits and events are not significantly polarized. According to the results of RFI environment tests at the FAST site, there are two major types of RFI sources within the $1.05-1.45$ GHz frequency band, civil aviation and navigation satellites \citep{2021RAA....21...18W}. We find only a small fraction of the hits falling within the civil aviation band ($1030 - 1140$ MHz). In contrast, hits within the navigation satellite bands ($1176.45 \pm 1.023$ MHz, $1207.14 \pm 2.046$ MHz, $1227.6 \pm 10$ MHz, $1246.0 - 1256.5$ MHz, $1268.52 \pm 10.23$ MHz, $1381.05 \pm 1.023$ MHz) account for a considerable proportion of the hits, implying that they are a major RFI source. 
%The proportions of the hits with positive, zero, and negative drift rates are 5.6\%, 65.8\% and 28.6\% for XX, and 5.4\%, 66.1\% and 28.5\% for YY.
Non-drift hits are in the majority as expected, since ground-based RFI sources are mostly stationary. The bias towards negative drift rate results from the downward relative acceleration vectors of non-geosynchronous satellites. The majority of the events have low S/N because weak RFI signals are likely to be detected below the S/N threshold by the reference beams but above the S/N threshold by Beam 1 accidentally, thus passing our event selection.

%\subsection{Signal Identification}

All the events selected by the program are re-examined by visual inspection of the dynamic time–frequency spectra (waterfall plots). we plot the dynamic spectra of the events using the BLIMPY package \citep{2019JOSS....4.1554P}, for beam 1 and reference  beams. We reject events when it is clear by eye that the  event was present in reference beams, but was not  detected above the turboSETI S/N $>$ 10 threshold. After visual inspection of these events, we found that all of them could be attributed to RFI. Most events are apparent false positives, they can be clearly seen in the reference beam.
21 events of XX and 28 events of YY, a total of 49 events were caused by instrument RFI \citep{2022AJ....164..160T}. These frequencies can be linearly combined by the nominal frequencies (33.3333 MHz, 125.00 MHz and 156.25 MHz) of the clock oscillators used by the Roach 2 FPGA board, so these events are attributed to the harmonics generated by the clock oscillators. So we attribute all the events to RFI.

\section{Discussion} \label{sec: Discussion}
\subsection{The minimum detectable EIRP from Barnard's star}

With the unprecedented sensitivity of FAST, we can thus compute the minimum detectable EIRP from Barnard's star, and it is defined as 

\begin{equation}
    \text{EIRP}_{\text{min}}=4\pi d^{2} S_{\text{min}},
\label{fuc:1}
\end{equation}
where $d$ is the distance to the target. The Gaia Data Release 3 \citep{2023A&A...674A...1G} shows the closest target in our observations, is 1.8 pc away. $S_{\text{min}}$ is the minimum detectable flux. 
For extremely narrowband signal detection (where the bandwidth of the transmitter signal is narrower or equal to the spectral resolution of observation), the minimum detectable flux $S_{\text{min}}$ is given by:
\begin{equation}
    S_{\text{min}} = \sigma_{\text{min}}~ \frac{2k_{\text{B}}T_{\text{sys}}}{A_{\text{eff}}} \sqrt{\frac{\delta\nu}{n_{\text{pol}} ~ t_{\text{obs}}}},
\label{fuc:2}    
\end{equation}
where $\sigma_{\text{min}}$ is the S/N threshold, $k_{\text{B}}$ is the Boltzmann constant, $T_{\text{sys}}$ is the system temperature and $A_{\text{eff}}$ is the effective collecting area. The ratio $A_{\text{eff}}/T_{\text{sys}}$ is also known as the sensitivity of the telescope. For the FAST L-band 19-beam receiver, the sensitivity is reported to be $\sim 2000$ m$^{2}$/K \citep{2011IJMPD..20..989N,2016RaSc...51.1060L,2019SCPMA..6259502J}. $\delta\nu$ is the frequency resolution, $n_{\text{pol}}$ is the polarization number, and $t_{\text{obs}}$ is the observation duration \citep{2017ApJ...849..104E}. Due to the separate search of data for XX and YY, the polarization number is 1. It is worth noting that this expression is different from the $S_{\text{min}}$ for observations aimed at astrophysical signals, where the signal bandwidth is wider than the frequency resolution, so the unit of $S_{\text{min}}$ here is W/m$^{2}$ rather than Jy ($10^{-26}$ W m$^{-2}$ Hz$^{-1}$). In this work, we observe the Barnard's star with a frequency resolution of $\sim 7.5$ Hz and a observation time of 1200s. The S/N threshold is set to 10.

We calculate $S_{\text{min}} = 1.09 \times 10^{-26}$ W/m$^{2}$ (1.09 Jy Hz) and $\text{EIRP}_{\text{min}} = 4.36 \times 10^{8}$ W.
For comparison, the EIRP of a typical Earth-based long-range radar transmitter is of the order of $\sim$10$^{9}$~W while a typical planetary radar is of the order of $\sim$10$^{13}$~W. 
Barnard's Star b is a planet in the Barnard's Star system \citep{2019ApJ...875L..12A}. Assuming that there is a radio transmitter on Barnard's Star b and sends signals to us with EIRP of $\sim$10$^{9}$~W and $\sim$10$^{13}$~W, we calculate the S/N of the signals observed by FAST to be in the order of $\sim 10$ and $\sim$10$^{5}$ respectively according to Function \ref{fuc:1} and Function \ref{fuc:2}. We simulated such hypothetical ETI signal (Figure \ref{fig:snr}) using setigen \citep{2022AJ....163..222B}.

\subsection{Drift rate estimation of a hypothetical transmitter on Barnard's star b}

\begin{figure}[htbp]
 \centering
 \includegraphics[width=0.8\textwidth]{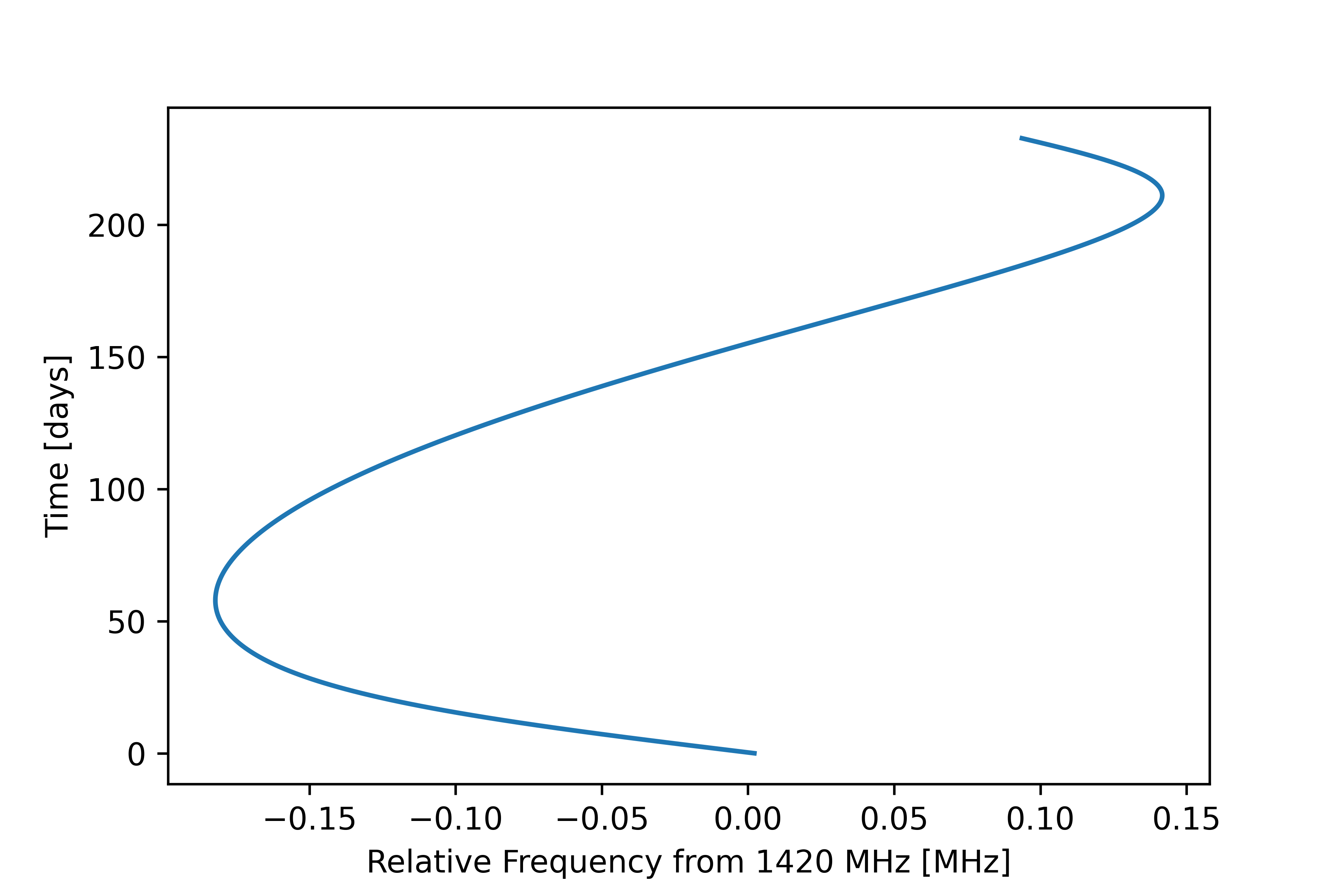}
 \caption{\label{fig:longterm} Pseudosinusoidal relative frequency curves of long-term observations for Barnard's Star b.}
 \end{figure}

If the duty cycle of the radio transmitter on the surface of Barnard's star b is 100$\%$, and the observation period can be long enough, the obtained ETI signal should fully or partially present a pseudo-sinusoidal curve. According to the parameters of Barnard's star b (Table \ref{table:parameters}) and Earth  \citep{2022ApJ...938....1L}, we simulated the long-term observation of Barnard's star, as shown in Figure \ref{fig:longterm}, the asymmetry of the curve is caused by the high eccentricities of the exoplanets' orbits \citep{2022ApJ...938....1L}. For the transmitter operating on the surface of Barnard's star b at 1420 MHz, the maximum drift rate we detected is 0.25 Hz/s.

FAST can easily detect weaker signals such as the communication signals we routinely generate from our long-range aircraft surveillance radars. Although factors such as whether extraterrestrial intelligence is sending a narrowband signal to Earth, and the duration of the transmission will make it impossible to predict the likelihood of detecting a signal. We can use the above frequency drift methods to verify a potential ETI signal of interest if we have enough observation time to carry out re-observation \citep{2022ApJ...938....1L}.
%difficult to detect the signal.

 \begin{figure}[htbp]
 \centering
 \includegraphics[width=0.8\textwidth]{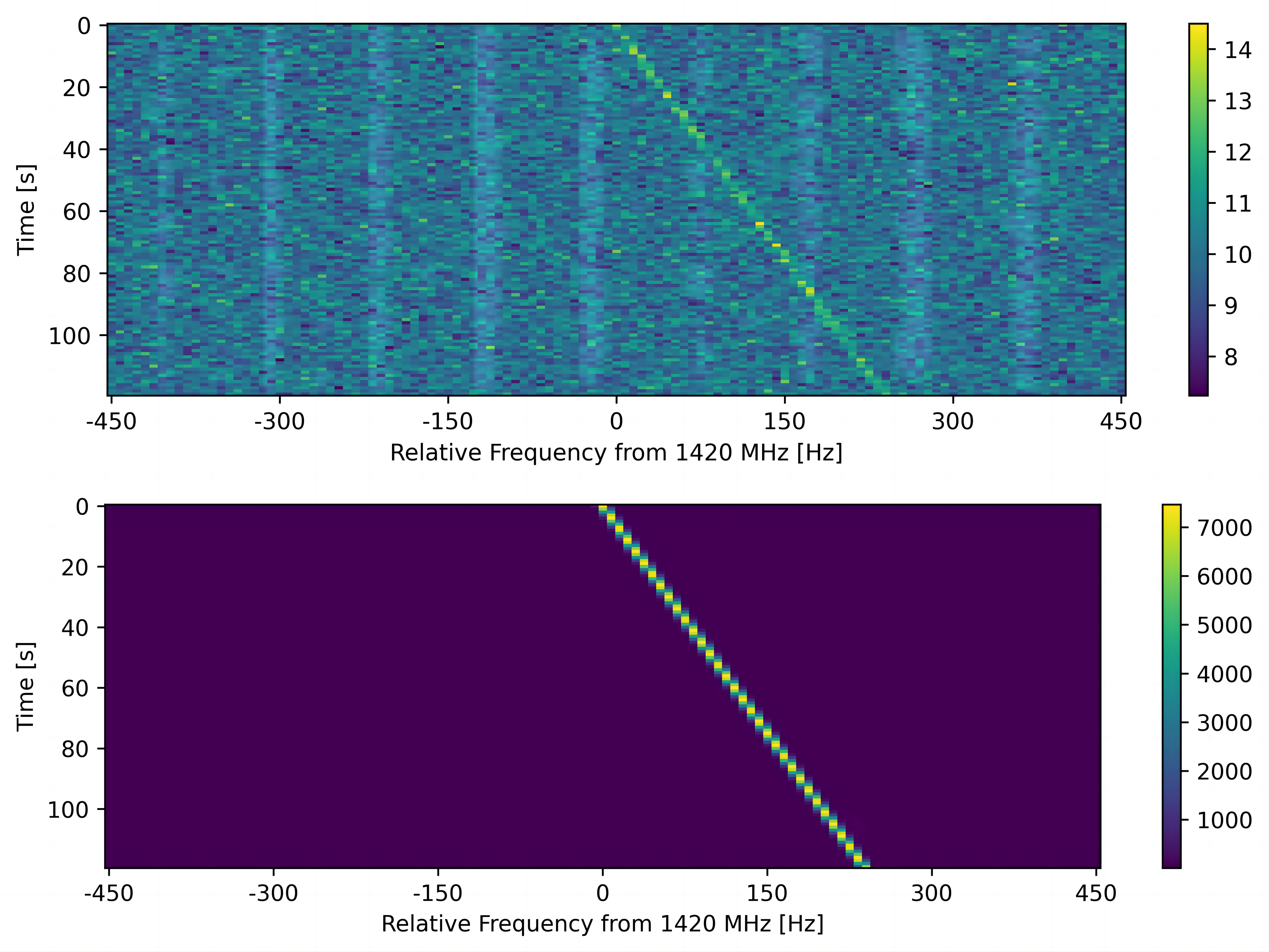}
 \caption{\label{fig:snr} Simulated FAST received signals from radio transmitters of different powers on the surface of Barnard's Star B (drift rate is 0.2 Hz/s): top (EIRP$\approx 10^{9}$~W), bottom (EIR$\approx 10^{13}$~W).}
 \end{figure}

\begin{table*}[!t]
	\caption{ Physical properties of Barnard’s star b}
	\centering
    \begin{threeparttable}
	\begin{tabular}{ll}
		\hline \hline Parameters & Value \\
		\hline Orbital period (d) & $232.80_{-0.41}^{+0.38} $ \\
	  Eccentricity & $0.32_{-0.15}^{+0.10} $ \\
        Star mass($M_{\odot}$) & $0.161\pm0.0036$\\
        Orbital semi-major axis(AU) & $0.404\pm0.018$\\
        R.A & $17^{\text{h}} 57^{\text{m}} 48.49847^{\text{s}}$\\
        Decl. & $+04^{\circ}41'36.1139''$\\
		\hline
	\end{tabular}
     \begin{tablenotes}
        \footnotesize
        \item \textbf{Notes.} Orbital period, Eccentricity and Orbital semi-major axis are from \cite{2018Natur.563..365R}. Star mass is from \cite{2021ApJ...918...40P}. R.A and Decl. are from \cite{2023A&A...674A...1G}. $M_{\odot}$ is the mass of the Sun. The rotation factor is ignored.
        %\item M: The mass of the Sun.
        %\item M: The mass of Earth.
        %\item $i$: The orbital inclination.
        %\item $M$: The true planetary mass.
        %\item ${\alpha}$:The true astrometric semi-amplitude.
        %\item BJD: Barycentric Julian date.
     \end{tablenotes}
    \end{threeparttable}
	\label{table:parameters}
\end{table*}

\section{Conclusions} \label{sec: Conclusions}
FAST is the world's largest single-antenna radio telescope, SETI is one of its five science missions. Its extremely high sensitivity in low-frequency radio L-band is of great significance to SETI.  Barnard's star is the second closest star system to the sun and the closest star in FAST sky region coverage. 
We conduct searches for narrow-band drifting radio signals toward Barnard's star with FAST. Recording the data on the SETI backend with a frequency resolution of $\sim$7.5 Hz, we search narrow-band signals across $1.05-1.45$ GHz with drift rates within $\pm 4$ Hz/s and S/Ns above 10, in two orthogonal linear polarization directions separately. 
%Following the Equation 7 in \cite{2022AJ....164..160T}
we calculate the minimum $\text{EIRP}_{\text{min}} = 4.36 \times 10^{8}$ W, which achieves an unprecedented sensitivity. This $\text{EIRP}_{\text{min}}$ is 7 times smaller than Green Bank Telescope observations of Barnard's star and 4.4 times smaller than Parkes Telescope observations of the closest star, Prox Cen \citep{2021NatAs.tmp..203S}.
Although, our observations find no solid evidence for 100\%-duty cycle radio transmitters emitting between 1.05 and 1.45 GHz with an EIRP above $4.36 \times 10^{8} ~\text{W}$.
Barnard's star remains an interesting target for technosignature searches, and in the future, we will continue to observe it and search for more types of technical features such as periodic signals \citep{2023AJ....165..255S}.
And we will conduct SETI surveys of more nearby stars as well as exoplanetary magnitude systems with FAST.

\vspace{12 pt}
%\begin{acknowledgments}
We sincerely appreciate the insightful and useful comments of referee and editor, which helped us improve our manuscript. We sincerely thank Vishal Gajjar for the kind and useful discussions.
T.-J.Z.(张同杰) dedicates this paper to the memory of his mother, Yu-Zhen Han(韩玉珍), who passed away 3 yr ago (2020 August 26).
This work was supported by the National SKA Program of China
(2022SKA0110202), the National Science Foundation of China
(grant No.11929301).
%, and the National Key R$\&$D Program of China (2017YFA0402600). 
This work is finished on the servers from FAST Data Center in Dezhou University. 
\bibliography{text}

\begin{thebibliography}{}
\expandafter\ifx\csname natexlab\endcsname\relax\def\natexlab#1{#1}\fi
\providecommand{\url}[1]{\href{#1}{#1}}
\providecommand{\dodoi}[1]{doi:~\href{http://doi.org/#1}{\nolinkurl{#1}}}
\providecommand{\doeprint}[1]{\href{http://ascl.net/#1}{\nolinkurl{http://ascl.net/#1}}}
\providecommand{\doarXiv}[1]{\href{https://arxiv.org/abs/#1}{\nolinkurl{https://arxiv.org/abs/#1}}}

\bibitem[{{Alvarado-G{\'o}mez} {et~al.}(2019){Alvarado-G{\'o}mez}, {Garraffo},
  {Drake}, {Brown}, {Oishi}, {Moschou}, \& {Cohen}}]{2019ApJ...875L..12A}
{Alvarado-G{\'o}mez}, J.~D., {Garraffo}, C., {Drake}, J.~J., {et~al.} 2019,
  \apjl, 875, L12, \dodoi{10.3847/2041-8213/ab1489}

\bibitem[{{Anglada-Escud{\'e}} {et~al.}(2016){Anglada-Escud{\'e}}, {Amado},
  {Barnes}, {Berdi{\~n}as}, {Butler}, {Coleman}, {de La Cueva}, {Dreizler},
  {Endl}, {Giesers}, {Jeffers}, {Jenkins}, {Jones}, {Kiraga}, {K{\"u}rster},
  {L{\'o}pez-Gonz{\'a}lez}, {Marvin}, {Morales}, {Morin}, {Nelson}, {Ortiz},
  {Ofir}, {Paardekooper}, {Reiners}, {Rodr{\'\i}guez},
  {Rodr{\'\i}guez-L{\'o}pez}, {Sarmiento}, {Strachan}, {Tsapras}, {Tuomi}, \&
  {Zechmeister}}]{2016Natur.536..437A}
{Anglada-Escud{\'e}}, G., {Amado}, P.~J., {Barnes}, J., {et~al.} 2016, \nat,
  536, 437, \dodoi{10.1038/nature19106}

\bibitem[{{Barnard}(1916)}]{1916AJ.....29..181B}
{Barnard}, E.~E. 1916, \aj, 29, 181, \dodoi{10.1086/104156}

\bibitem[{{Brzycki} {et~al.}(2022){Brzycki}, {Siemion}, {Pater}, {Croft},
  {Hoang}, {Ng}, {Price}, {Sheikh}, \& {Zheng}}]{2022AJ....163..222B}
{Brzycki}, B., {Siemion}, A. P.~V., {Pater}, I.~d., {et~al.} 2022, \aj, 163,
  222, \dodoi{10.3847/1538-3881/ac5e3d}

\bibitem[{{Cordes} {et~al.}(1997){Cordes}, {Lazio}, \&
  {Sagan}}]{1997ApJ...487..782C}
{Cordes}, J.~M., {Lazio}, J.~W., \& {Sagan}, C. 1997, \apj, 487, 782,
  \dodoi{10.1086/304620}

\bibitem[{{Drake}(1961)}]{1961PhT....14d..40D}
{Drake}, F.~D. 1961, Physics Today, 14, 40, \dodoi{10.1063/1.3057500}

\bibitem[{{Enriquez} \& {Price}(2019)}]{2019ascl.soft06006E}
{Enriquez}, E., \& {Price}, D. 2019, {turboSETI: Python-based SETI search
  algorithm}.
\newblock \doeprint{1906.006}

\bibitem[{{Enriquez} {et~al.}(2017){Enriquez}, {Siemion}, {Foster}, {Gajjar},
  {Hellbourg}, {Hickish}, {Isaacson}, {Price}, {Croft}, {DeBoer}, {Lebofsky},
  {MacMahon}, \& {Werthimer}}]{2017ApJ...849..104E}
{Enriquez}, J.~E., {Siemion}, A., {Foster}, G., {et~al.} 2017, \apj, 849, 104,
  \dodoi{10.3847/1538-4357/aa8d1b}

\bibitem[{{Gaia Collaboration} {et~al.}(2023){Gaia Collaboration}, {Vallenari},
  {Brown}, {Prusti}, {de Bruijne}, {Arenou}, {Babusiaux}, {Biermann},
  {Creevey}, {Ducourant}, {Evans}, {Eyer}, {Guerra}, {Hutton}, {Jordi},
  {Klioner}, {Lammers}, {Lindegren}, {Luri}, {Mignard}, {Panem}, {Pourbaix},
  {Randich}, {Sartoretti}, {Soubiran}, {Tanga}, {Walton}, {Bailer-Jones},
  {Bastian}, {Drimmel}, {Jansen}, {Katz}, {Lattanzi}, {van Leeuwen}, {Bakker},
  {Cacciari}, {Casta{\~n}eda}, {De Angeli}, {Fabricius}, {Fouesneau},
  {Fr{\'e}mat}, {Galluccio}, {Guerrier}, {Heiter}, {Masana}, {Messineo},
  {Mowlavi}, {Nicolas}, {Nienartowicz}, {Pailler}, {Panuzzo}, {Riclet}, {Roux},
  {Seabroke}, {Sordo}, {Th{\'e}venin}, {Gracia-Abril}, {Portell}, {Teyssier},
  {Altmann}, {Andrae}, {Audard}, {Bellas-Velidis}, {Benson}, {Berthier},
  {Blomme}, {Burgess}, {Busonero}, {Busso}, {C{\'a}novas}, {Carry}, {Cellino},
  {Cheek}, {Clementini}, {Damerdji}, {Davidson}, {de Teodoro}, {Nu{\~n}ez
  Campos}, {Delchambre}, {Dell'Oro}, {Esquej}, {Fern{\'a}ndez-Hern{\'a}ndez},
  {Fraile}, {Garabato}, {Garc{\'\i}a-Lario}, {Gosset}, {Haigron}, {Halbwachs},
  {Hambly}, {Harrison}, {Hern{\'a}ndez}, {Hestroffer}, {Hodgkin}, {Holl},
  {Jan{\ss}en}, {Jevardat de Fombelle}, {Jordan}, {Krone-Martins}, {Lanzafame},
  {L{\"o}ffler}, {Marchal}, {Marrese}, {Moitinho}, {Muinonen}, {Osborne},
  {Pancino}, {Pauwels}, {Recio-Blanco}, {Reyl{\'e}}, {Riello}, {Rimoldini},
  {Roegiers}, {Rybizki}, {Sarro}, {Siopis}, {Smith}, {Sozzetti}, {Utrilla},
  {van Leeuwen}, {Abbas}, {{\'A}brah{\'a}m}, {Abreu Aramburu}, {Aerts},
  {Aguado}, {Ajaj}, {Aldea-Montero}, {Altavilla}, {{\'A}lvarez}, {Alves},
  {Anders}, {Anderson}, {Anglada Varela}, {Antoja}, {Baines}, {Baker},
  {Balaguer-N{\'u}{\~n}ez}, {Balbinot}, {Balog}, {Barache}, {Barbato},
  {Barros}, {Barstow}, {Bartolom{\'e}}, {Bassilana}, {Bauchet}, {Becciani},
  {Bellazzini}, {Berihuete}, {Bernet}, {Bertone}, {Bianchi}, {Binnenfeld},
  {Blanco-Cuaresma}, {Blazere}, {Boch}, {Bombrun}, {Bossini}, {Bouquillon},
  {Bragaglia}, {Bramante}, {Breedt}, {Bressan}, {Brouillet}, {Brugaletta},
  {Bucciarelli}, {Burlacu}, {Butkevich}, {Buzzi}, {Caffau}, {Cancelliere},
  {Cantat-Gaudin}, {Carballo}, {Carlucci}, {Carnerero}, {Carrasco},
  {Casamiquela}, {Castellani}, {Castro-Ginard}, {Chaoul}, {Charlot}, {Chemin},
  {Chiaramida}, {Chiavassa}, {Chornay}, {Comoretto}, {Contursi}, {Cooper},
  {Cornez}, {Cowell}, {Crifo}, {Cropper}, {Crosta}, {Crowley}, {Dafonte},
  {Dapergolas}, {David}, {David}, {de Laverny}, {De Luise}, {De March}, {De
  Ridder}, {de Souza}, {de Torres}, {del Peloso}, {del Pozo}, {Delbo},
  {Delgado}, {Delisle}, {Demouchy}, {Dharmawardena}, {Di Matteo}, {Diakite},
  {Diener}, {Distefano}, {Dolding}, {Edvardsson}, {Enke}, {Fabre}, {Fabrizio},
  {Faigler}, {Fedorets}, {Fernique}, {Fienga}, {Figueras}, {Fournier},
  {Fouron}, {Fragkoudi}, {Gai}, {Garcia-Gutierrez}, {Garcia-Reinaldos},
  {Garc{\'\i}a-Torres}, {Garofalo}, {Gavel}, {Gavras}, {Gerlach}, {Geyer},
  {Giacobbe}, {Gilmore}, {Girona}, {Giuffrida}, {Gomel}, {Gomez},
  {Gonz{\'a}lez-N{\'u}{\~n}ez}, {Gonz{\'a}lez-Santamar{\'\i}a},
  {Gonz{\'a}lez-Vidal}, {Granvik}, {Guillout}, {Guiraud},
  {Guti{\'e}rrez-S{\'a}nchez}, {Guy}, {Hatzidimitriou}, {Hauser}, {Haywood},
  {Helmer}, {Helmi}, {Sarmiento}, {Hidalgo}, {Hilger}, {H{\l}adczuk}, {Hobbs},
  {Holland}, {Huckle}, {Jardine}, {Jasniewicz}, {Jean-Antoine Piccolo},
  {Jim{\'e}nez-Arranz}, {Jorissen}, {Juaristi Campillo}, {Julbe}, {Karbevska},
  {Kervella}, {Khanna}, {Kontizas}, {Kordopatis}, {Korn}, {K{\'o}sp{\'a}l},
  {Kostrzewa-Rutkowska}, {Kruszy{\'n}ska}, {Kun}, {Laizeau}, {Lambert},
  {Lanza}, {Lasne}, {Le Campion}, {Lebreton}, {Lebzelter}, {Leccia}, {Leclerc},
  {Lecoeur-Taibi}, {Liao}, {Licata}, {Lindstr{\o}m}, {Lister}, {Livanou},
  {Lobel}, {Lorca}, {Loup}, {Madrero Pardo}, {Magdaleno Romeo}, {Managau},
  {Mann}, {Manteiga}, {Marchant}, {Marconi}, {Marcos}, {Marcos Santos},
  {Mar{\'\i}n Pina}, {Marinoni}, {Marocco}, {Marshall}, {Martin Polo},
  {Mart{\'\i}n-Fleitas}, {Marton}, {Mary}, {Masip}, {Massari},
  {Mastrobuono-Battisti}, {Mazeh}, {McMillan}, {Messina}, {Michalik}, {Millar},
  {Mints}, {Molina}, {Molinaro}, {Moln{\'a}r}, {Monari}, {Mongui{\'o}},
  {Montegriffo}, {Montero}, {Mor}, {Mora}, {Morbidelli}, {Morel}, {Morris},
  {Muraveva}, {Murphy}, {Musella}, {Nagy}, {Noval}, {Oca{\~n}a}, {Ogden},
  {Ordenovic}, {Osinde}, {Pagani}, {Pagano}, {Palaversa}, {Palicio},
  {Pallas-Quintela}, {Panahi}, {Payne-Wardenaar}, {Pe{\~n}alosa Esteller},
  {Penttil{\"a}}, {Pichon}, {Piersimoni}, {Pineau}, {Plachy}, {Plum}, {Poggio},
  {Pr{\v{s}}a}, {Pulone}, {Racero}, {Ragaini}, {Rainer}, {Raiteri}, {Rambaux},
  {Ramos}, {Ramos-Lerate}, {Re Fiorentin}, {Regibo}, {Richards}, {Rios Diaz},
  {Ripepi}, {Riva}, {Rix}, {Rixon}, {Robichon}, {Robin}, {Robin}, {Roelens},
  {Rogues}, {Rohrbasser}, {Romero-G{\'o}mez}, {Rowell}, {Royer}, {Ruz Mieres},
  {Rybicki}, {Sadowski}, {S{\'a}ez N{\'u}{\~n}ez}, {Sagrist{\`a} Sell{\'e}s},
  {Sahlmann}, {Salguero}, {Samaras}, {Sanchez Gimenez}, {Sanna},
  {Santove{\~n}a}, {Sarasso}, {Schultheis}, {Sciacca}, {Segol}, {Segovia},
  {S{\'e}gransan}, {Semeux}, {Shahaf}, {Siddiqui}, {Siebert}, {Siltala},
  {Silvelo}, {Slezak}, {Slezak}, {Smart}, {Snaith}, {Solano}, {Solitro},
  {Souami}, {Souchay}, {Spagna}, {Spina}, {Spoto}, {Steele},
  {Steidelm{\"u}ller}, {Stephenson}, {S{\"u}veges}, {Surdej}, {Szabados},
  {Szegedi-Elek}, {Taris}, {Taylor}, {Teixeira}, {Tolomei}, {Tonello}, {Torra},
  {Torra}, {Torralba Elipe}, {Trabucchi}, {Tsounis}, {Turon}, {Ulla}, {Unger},
  {Vaillant}, {van Dillen}, {van Reeven}, {Vanel}, {Vecchiato}, {Viala},
  {Vicente}, {Voutsinas}, {Weiler}, {Wevers}, {Wyrzykowski}, {Yoldas}, {Yvard},
  {Zhao}, {Zorec}, {Zucker}, \& {Zwitter}}]{2023A&A...674A...1G}
{Gaia Collaboration}, {Vallenari}, A., {Brown}, A.~G.~A., {et~al.} 2023, \aap,
  674, A1, \dodoi{10.1051/0004-6361/202243940}

\bibitem[{{Gajjar} {et~al.}(2021){Gajjar}, {Perez}, {Siemion}, {Foster},
  {Brzycki}, {Chatterjee}, {Chen}, {Cordes}, {Croft}, {Czech}, {DeBoer},
  {DeMarines}, {Drew}, {Gowanlock}, {Isaacson}, {Lacki}, {Lebofsky},
  {MacMahon}, {Morrison}, {Ng}, {de Pater}, {Price}, {Sheikh}, {Suresh},
  {Webb}, \& {Pete Worden}}]{2021AJ....162...33G}
{Gajjar}, V., {Perez}, K.~I., {Siemion}, A. P.~V., {et~al.} 2021, \aj, 162, 33,
  \dodoi{10.3847/1538-3881/abfd36}

\bibitem[{{Harp} {et~al.}(2016){Harp}, {Richards}, {Tarter}, {Dreher},
  {Jordan}, {Shostak}, {Smolek}, {Kilsdonk}, {Wilcox}, {Wimberly}, {Ross},
  {Barott}, {Ackermann}, \& {Blair}}]{2016AJ....152..181H}
{Harp}, G.~R., {Richards}, J., {Tarter}, J.~C., {et~al.} 2016, \aj, 152, 181,
  \dodoi{10.3847/0004-6256/152/6/181}

\bibitem[{{Jiang} {et~al.}(2019){Jiang}, {Yue}, {Gan}, {Yao}, {Li}, {Pan},
  {Sun}, {Yu}, {Liu}, {Tang}, {Qian}, {Lu}, {Yan}, {Peng}, {Zhang}, {Wang},
  {Li}, \& {Li}}]{2019SCPMA..6259502J}
{Jiang}, P., {Yue}, Y., {Gan}, H., {et~al.} 2019, Science China Physics,
  Mechanics, and Astronomy, 62, 959502, \dodoi{10.1007/s11433-018-9376-1}

\bibitem[{{Li} \& {Pan}(2016)}]{2016RaSc...51.1060L}
{Li}, D., \& {Pan}, Z. 2016, Radio Science, 51, 1060,
  \dodoi{10.1002/2015RS005877}

\bibitem[{{Li} {et~al.}(2022){Li}, {Zhao}, {Tao}, {Zhang}, \&
  {Xiao-Hui}}]{2022ApJ...938....1L}
{Li}, J.-K., {Zhao}, H.-C., {Tao}, Z.-Z., {Zhang}, T.-J., \& {Xiao-Hui}, S.
  2022, \apj, 938, 1, \dodoi{10.3847/1538-4357/ac90bd}

\bibitem[{{Luan} {et~al.}(2023){Luan}, {Tao}, {Zhao}, {Huang}, {Li}, {Liu},
  {Wang}, {Liu}, {Zhang}, {Gajjar}, \& {Werthimer}}]{2023AJ....165..132L}
{Luan}, X.-H., {Tao}, Z.-Z., {Zhao}, H.-C., {et~al.} 2023, \aj, 165, 132,
  \dodoi{10.3847/1538-3881/acb706}

\bibitem[{{Lubin} {et~al.}(2021){Lubin}, {Robertson}, {Stefansson}, {Ninan},
  {Mahadevan}, {Endl}, {Ford}, {Wright}, {Beard}, {Bender}, {Cochran},
  {Diddams}, {Fredrick}, {Halverson}, {Kanodia}, {Metcalf}, {Ramsey}, {Roy},
  {Schwab}, \& {Terrien}}]{2021AJ....162...61L}
{Lubin}, J., {Robertson}, P., {Stefansson}, G., {et~al.} 2021, \aj, 162, 61,
  \dodoi{10.3847/1538-3881/ac0057}

\bibitem[{{Nan} {et~al.}(2011){Nan}, {Li}, {Jin}, {Wang}, {Zhu}, {Zhu},
  {Zhang}, {Yue}, \& {Qian}}]{2011IJMPD..20..989N}
{Nan}, R., {Li}, D., {Jin}, C., {et~al.} 2011, International Journal of Modern
  Physics D, 20, 989, \dodoi{10.1142/S0218271811019335}

\bibitem[{{Pinchuk} {et~al.}(2019){Pinchuk}, {Margot}, {Greenberg}, {Ayalde},
  {Bloxham}, {Boddu}, {Gerardo Chinchilla-Garcia}, {Cliffe}, {Gallagher},
  {Hart}, {Hesford}, {Mizrahi}, {Pike}, {Rodger}, {Sayki}, {Schneck}, {Tan},
  {{\textquotedblleft}Yolanda{\textquotedblright} Xiao}, \&
  {Lynch}}]{2019AJ....157..122P}
{Pinchuk}, P., {Margot}, J.-L., {Greenberg}, A.~H., {et~al.} 2019, \aj, 157,
  122, \dodoi{10.3847/1538-3881/ab0105}

\bibitem[{{Pineda} {et~al.}(2021){Pineda}, {Youngblood}, \&
  {France}}]{2021ApJ...918...40P}
{Pineda}, J.~S., {Youngblood}, A., \& {France}, K. 2021, \apj, 918, 40,
  \dodoi{10.3847/1538-4357/ac0aea}

\bibitem[{{Price} {et~al.}(2019){Price}, {Enriquez}, {Chen}, \&
  {Siebert}}]{2019JOSS....4.1554P}
{Price}, D., {Enriquez}, J., {Chen}, Y., \& {Siebert}, M. 2019, The Journal of
  Open Source Software, 4, 1554, \dodoi{10.21105/joss.01554}

\bibitem[{{Price} {et~al.}(2020){Price}, {Enriquez}, {Brzycki}, {Croft},
  {Czech}, {DeBoer}, {DeMarines}, {Foster}, {Gajjar}, {Gizani}, {Hellbourg},
  {Isaacson}, {Lacki}, {Lebofsky}, {MacMahon}, {Pater}, {Siemion}, {Werthimer},
  {Green}, {Kaczmarek}, {Maddalena}, {Mader}, {Drew}, \&
  {Worden}}]{2020AJ....159...86P}
{Price}, D.~C., {Enriquez}, J.~E., {Brzycki}, B., {et~al.} 2020, \aj, 159, 86,
  \dodoi{10.3847/1538-3881/ab65f1}

\bibitem[{{Ribas} {et~al.}(2018){Ribas}, {Tuomi}, {Reiners}, {Butler},
  {Morales}, {Perger}, {Dreizler}, {Rodr{\'\i}guez-L{\'o}pez}, {Gonz{\'a}lez
  Hern{\'a}ndez}, {Rosich}, {Feng}, {Trifonov}, {Vogt}, {Caballero}, {Hatzes},
  {Herrero}, {Jeffers}, {Lafarga}, {Murgas}, {Nelson}, {Rodr{\'\i}guez},
  {Strachan}, {Tal-Or}, {Teske}, {Toledo-Padr{\'o}n}, {Zechmeister},
  {Quirrenbach}, {Amado}, {Azzaro}, {B{\'e}jar}, {Barnes}, {Berdi{\~n}as},
  {Burt}, {Coleman}, {Cort{\'e}s-Contreras}, {Crane}, {Engle}, {Guinan},
  {Haswell}, {Henning}, {Holden}, {Jenkins}, {Jones}, {Kaminski}, {Kiraga},
  {K{\"u}rster}, {Lee}, {L{\'o}pez-Gonz{\'a}lez}, {Montes}, {Morin}, {Ofir},
  {Pall{\'e}}, {Rebolo}, {Reffert}, {Schweitzer}, {Seifert}, {Shectman},
  {Staab}, {Street}, {Su{\'a}rez Mascare{\~n}o}, {Tsapras}, {Wang}, \&
  {Anglada-Escud{\'e}}}]{2018Natur.563..365R}
{Ribas}, I., {Tuomi}, M., {Reiners}, A., {et~al.} 2018, \nat, 563, 365,
  \dodoi{10.1038/s41586-018-0677-y}

\bibitem[{{Seager}(2013)}]{2013Sci...340..577S}
{Seager}, S. 2013, Science, 340, 577, \dodoi{10.1126/science.1232226}

\bibitem[{{Sheikh} {et~al.}(2020){Sheikh}, {Siemion}, {Enriquez}, {Price},
  {Isaacson}, {Lebofsky}, {Gajjar}, \& {Kalas}}]{2020AJ....160...29S}
{Sheikh}, S.~Z., {Siemion}, A., {Enriquez}, J.~E., {et~al.} 2020, \aj, 160, 29,
  \dodoi{10.3847/1538-3881/ab9361}

\bibitem[{{Shostak}(2011)}]{2011AcAau..68..347S}
{Shostak}, S. 2011, Acta Astronautica, 68, 347,
  \dodoi{10.1016/j.actaastro.2009.07.013}

\bibitem[{{Siemion} {et~al.}(2013){Siemion}, {Demorest}, {Korpela},
  {Maddalena}, {Werthimer}, {Cobb}, {Howard}, {Langston}, {Lebofsky}, {Marcy},
  \& {Tarter}}]{2013ApJ...767...94S}
{Siemion}, A. P.~V., {Demorest}, P., {Korpela}, E., {et~al.} 2013, \apj, 767,
  94, \dodoi{10.1088/0004-637X/767/1/94}

\bibitem[{{Smith} {et~al.}(2021){Smith}, {Price}, {Sheikh}, {Czech}, {Croft},
  {DeBoer}, {Gajjar}, {Isaacson}, {Lacki}, {Lebofsky}, {MacMahon}, {Ng},
  {Perez}, {Siemion}, {Webb}, {Drew}, {Worden}, \& {Zic}}]{2021NatAs.tmp..203S}
{Smith}, S., {Price}, D.~C., {Sheikh}, S.~Z., {et~al.} 2021, Nature Astronomy,
  \dodoi{10.1038/s41550-021-01479-w}

\bibitem[{{Suresh} {et~al.}(2023){Suresh}, {Gajjar}, {Nagarajan}, {Sheikh},
  {Siemion}, {Lebofsky}, {MacMahon}, {Price}, \& {Croft}}]{2023AJ....165..255S}
{Suresh}, A., {Gajjar}, V., {Nagarajan}, P., {et~al.} 2023, \aj, 165, 255,
  \dodoi{10.3847/1538-3881/acccf0}

\bibitem[{{Tao} {et~al.}(2022){Tao}, {Zhao}, {Zhang}, {Gajjar}, {Zhu}, {Yue},
  {Zhang}, {Liu}, {Li}, {Zhang}, {Liu}, {Wang}, {Duan}, {Qian}, {Jin}, {Li},
  {Siemion}, {Jiang}, {Werthimer}, {Cobb}, {Korpela}, \&
  {Anderson}}]{2022AJ....164..160T}
{Tao}, Z.-Z., {Zhao}, H.-C., {Zhang}, T.-J., {et~al.} 2022, \aj, 164, 160,
  \dodoi{10.3847/1538-3881/ac8bd5}

\bibitem[{{Traas} {et~al.}(2021){Traas}, {Croft}, {Gajjar}, {Isaacson},
  {Lebofsky}, {MacMahon}, {Perez}, {Price}, {Sheikh}, {Siemion}, {Smith},
  {Drew}, \& {Worden}}]{2021AJ....161..286T}
{Traas}, R., {Croft}, S., {Gajjar}, V., {et~al.} 2021, \aj, 161, 286,
  \dodoi{10.3847/1538-3881/abf649}

\bibitem[{{Wang} {et~al.}(2021){Wang}, {Zhang}, {Hu}, {Huang}, {Zhu}, {Zhi},
  {Zhang}, {Fan}, \& {Yang}}]{2021RAA....21...18W}
{Wang}, Y., {Zhang}, H.-Y., {Hu}, H., {et~al.} 2021, Research in Astronomy and
  Astrophysics, 21, 018, \dodoi{10.1088/1674-4527/21/1/18}

\bibitem[{{Wells} {et~al.}(1981){Wells}, {Greisen}, \&
  {Harten}}]{1981A&AS...44..363W}
{Wells}, D.~C., {Greisen}, E.~W., \& {Harten}, R.~H. 1981, \aaps, 44, 363

\bibitem[{{Zhang} {et~al.}(2020){Zhang}, {Werthimer}, {Zhang}, {Cobb},
  {Korpela}, {Anderson}, {Gajjar}, {Lee}, {Li}, {Pei}, {Zhang}, {Huang},
  {Wang}, {Zhu}, {Duan}, {Zhang}, {Jin}, {Zhu}, \& {Li}}]{2020ApJ...891..174Z}
{Zhang}, Z.-S., {Werthimer}, D., {Zhang}, T.-J., {et~al.} 2020, \apj, 891, 174,
  \dodoi{10.3847/1538-4357/ab7376}

\bibitem[{{Zhu} {et~al.}(2020){Zhu}, {Li}, {Luo}, {Miao}, {Zhang}, {Spitler},
  {Lorimer}, {Kramer}, {Champion}, {Yue}, {Cameron}, {Cruces}, {Duan}, {Feng},
  {Han}, {Hobbs}, {Niu}, {Niu}, {Pan}, {Qian}, {Shi}, {Tang}, {Wang}, {Wang},
  {Yuan}, {Zhang}, {Zhang}, {Cao}, {Feng}, {Gan}, {Gao}, {Gu}, {Guo}, {Hao},
  {Huang}, {Huang}, {Jiang}, {Jin}, {Li}, {Li}, {Li}, {Liu}, {Pan}, {Peng},
  {Qian}, {Shi}, {Song}, {Song}, {Sun}, {Sun}, {Wang}, {Wang}, {Wang}, {Xie},
  {Yan}, {Yang}, {Yang}, {Yao}, {Yu}, {Yu}, {Zhang}, {Zhang}, {Zhang}, {Zheng},
  {Zhou}, {Zhu}, {Zhu}, {Zhu}, {Zhu}, \& {Zhu}}]{2020ApJ...895L...6Z}
{Zhu}, W., {Li}, D., {Luo}, R., {et~al.} 2020, \apjl, 895, L6,
  \dodoi{10.3847/2041-8213/ab8e46}

\end{thebibliography}


\end{CJK*}
\end{document}